\documentclass[11pt,a4paper]{article}
\usepackage{jheppub}
\usepackage{amsmath}
\usepackage[most]{tcolorbox}
\usepackage{dsfont}
\usepackage{ulem}
\usepackage{natbib}
\usepackage{xcolor}
\usepackage[hang,flushmargin]{footmisc}
\usepackage{tikz-cd}
\usepackage{enumitem}
\usepackage{amsfonts,amssymb, amscd,amsmath,latexsym,amsbsy,bm}
\usepackage{stmaryrd}
\usepackage{todonotes}
\usepackage{stackrel}
\usepackage{amsthm}

\usepackage{graphicx}
\usepackage{verbatim}

\title{\boldmath Ordinary limits of the hyperbolic hypergeometric integral identities}

\author{Mustafa Mullahasanoglu$^{a,b}$, Ali Mert T. Yetkin$^{a}$ and Reyhan Yumuşak$^{a}$}
\affiliation{
$^a$ Department of Physics, Bogazici University,\\ 34342 Bebek, Istanbul, Türkiye\\[-0.4cm]

$^b$ Feza Gursey Center for Physics and Mathematics, Bogazici University,\\ 34684, Kandilli,
Istanbul, Türkiye
}

\emailAdd{mustafa.mullahasanoglu@std.bogazici.edu.tr}
\emailAdd{ali.yetkin@std.bogazici.edu.tr}
\emailAdd{ reyhan.yumusak@std.bogazici.edu.tr}

\abstract{The computation of the partition function of supersymmetric gauge theories on compact manifolds can be reduced to matrix integrals by using the supersymmetric localization technique. Such matrix integrals in the case of three-dimensional supersymmetric gauge theories on lens space can be expressed in terms of hyperbolic hypergeometric integrals. By studying partition functions of supersymmetric dual theories, one can obtain new complicated identities for this type of special functions. We derive new ordinary hypergeometric identities from the reduction of certain hyperbolic hypergeometric integral identities obtained via supersymmetric infrared dualities.}

\begin{document}

	\maketitle

\section{Introduction}

Recent advances in partition function calculations for supersymmetric field theories have important mathematical implications.  They are often expressed in terms of hypergeometric integrals, which are of interest in both mathematics and physics.  One can obtain interesting integral identities for the supersymmetric dualities by equating the partition functions of the dual theories.

Here we consider the lens partition functions of three-dimensional supersymmetric theories, which can be written as hyperbolic hypergeometric integrals. We study reductions of hyperbolic hypergeometric integral identities to the ordinary hypergeometric ones and find known and some new hypergeometric integral identities. Since a two-sphere partition function can be expressed in terms of ordinary hypergeometric integrals, one can use these integral identities in the search for new supersymmetric dualities.

\section{Supersymmetric partition functions}

We start by defining the general form of the three-dimensional $\mathcal N=2$ partition function on the squashed lens space\footnote{	Recall that the lens space $S_b^3/\mathbb Z_r$ can be obtained from the squashed three-sphere 
	\begin{equation}
	S_b^3= \{(x,y)\in\mathbb{C}^2, b^2|x|^2+b^{-2}|y|^2=1\} \:,
	\end{equation}
	by making the identification $(x,y)\sim(e^{\frac{2\pi i}{r}}x,e^{\frac{2\pi i}{r}}y)$.} $S^3_b/\mathbb{Z}_r$.  The lens partition function\footnote{We refer the reader to \cite{Imamura:2012rq,Imamura:2013qxa,Gahramanov:2016ilb} for more details.} can be computed by a dimensional reduction of the four-dimensional lens superconformal index \cite{Benini:2011nc,Yamazaki:2013fva,Eren:2019ibl} and by the localization technique \cite{Imamura:2012rq,Imamura:2013qxa}. The partition function on the lens space can be reduced to the following matrix model
	\begin{equation}
	Z=\sum_y\int \frac{1}{|W|}\prod_j^{rank G} \frac{dz_j}{2\pi i r}Z_{\text{classical}}[z,y]Z_{\text{one-loop}}[z,y] \;.
	\end{equation}
	
The one-loop contribution can be written in terms of hyperbolic hypergeometric function\footnote{The reflection property of the lens hyperbolic gamma function is 
\begin{align}
\gamma_h(\omega_1+\omega_2-z,r-y;\omega_1,\omega_2)\gamma_h(z,y;\omega_1,\omega_2) 
=    1\:.
\end{align}
The following shorthand notations are used in the rest of the paper
\begin{align}
  \gamma_h(z,y)&=\gamma_h( z,y ;\omega_1,\omega_2)\:,
  \nonumber \\
  \gamma_h(\pm z,\pm y )&=\gamma_h( z,y )\gamma_h(- z,-y ) \:.
\end{align}} 
	\begin{align}
    \gamma_h(z,y;\omega_1,\omega_2) = \gamma^{(2)}&(-iz-i\omega_1y;-i\omega_1r,-i(\omega_1+\omega_2)) \nonumber \\ \times & \gamma^{(2)}(-iz -i\omega_2(r-y);-i\omega_2r,-i(\omega_1+\omega_2))\:,
    \label{lenshyperbolic}
\end{align}
with $\text{Im}(\omega_1/\omega_2)>0$ and the parameters take values as $r\in\{1,2,...\}$, $y\in \{0,1,...,r-1\}$. 

It is possible to reduce the supersymmetric partition functions of three-dimensional theories to the partition functions of two-dimensional supersymmetric gauge theories. This reduction for the dual theories leads to the equality of the partition functions expressed in terms of the ordinary hypergeometric integral identities. 

The supersymmetric partition function on $S^2$ was obtained in 
\cite{Benini:2012ui,Doroud:2012xw}. In this case, the one-loop contribution to the partition function can be written in terms of the ratio of the Euler $\Gamma(z)$ functions\footnote{
The reflection property of the Euler gamma function is  $$\Gamma(z)\Gamma(1-z)=\pi/\sin\pi z$$ which leads to the property
\begin{equation}
{\bf \Gamma}(z,-y)=(-1)^y{\bf \Gamma}(z,y)\:,
\label{gamma}\end{equation}
and we also have
\begin{equation}
{\bf \Gamma}(z,y){\bf \Gamma}(2-z,y)=1\:.
\end{equation}
Then we can write the reflection property as
\begin{equation}
{\bf \Gamma}(z,y){\bf \Gamma}(2-z,-y)=(-1)^y\:.
\end{equation}
} which is the usual Euler's gamma function
\begin{align}
    \boldsymbol{\Gamma}(z,y)=\frac{\Gamma\left(\frac{z+y}{2}\right)}{\Gamma\left(1-\frac{z-y}{2}\right)}
\end{align}

The hyperbolic gamma function reduces to the Euler gamma function by the asymptotic property 
\begin{equation}
	\lim_{\omega_2\to\infty} \Big(\frac{\omega_2}{2\pi\omega_1}\Big)^{\frac{z}{\omega_2}-\frac{1}{2}}\gamma^{(2)}(z;\omega_1,\omega_2)=\frac{\Gamma(z/\omega_1)}{\sqrt{2\pi}}\:.
\label{gamma_limit}
\end{equation}
By making the substitution \( z \to \omega_1 z \) in equation (\ref{lenshyperbolic}) and imposing the identification \( \omega_1 \leftrightarrow \omega_2 \), the asymptotic properties of the hyperbolic gamma function yield the following formula
\begin{equation}
	\lim_{r\to\infty} \gamma_h(z,y;\omega_1,\omega_2)=\Big(\frac{r}{4\pi}\Big)^{\frac{2-2z}{r}}\frac{\Gamma\left(\frac{z+y}{2}\right)}{\Gamma\left(1-\frac{z-y}{2}\right)}\equiv \Big(\frac{r}{4\pi}\Big)^{\frac{2-2z}{r}}{\bf \Gamma}(z,y)\:.
\end{equation}

\section{Integral identities}

\subsection{The integral identity I}

First we consider the following three-dimensional $\mathcal N=2$ supersymmetric duality \cite{Gahramanov:2015cva,Gahramanov:2016wxi}:

\textbf{Theory A.} the supersymmetric QCD with $SU(2)$ gauge group and with $SU(6)$ flavor group, chiral multiplets are in the fundamental representation of the gauge group and the flavor group

\textbf{Theory B.} has no gauge degrees of freedom, fifteen chiral multiplets of the theory are in the totally antisymmetric tensor representation of the flavor group.

The equality of the lens partition functions for this duality has the form of the following hyperbolic hypergeometric integral identity \cite{Gahramanov:2016ilb}
\begin{align}
 \sum_{y= 0}^{\lfloor \frac{r}{2} \rfloor}\epsilon(y)\int_{-\infty}^{\infty}\frac{dz}{2r\sqrt{-\omega_1\omega_2}}\prod_{i=1}^{6}\frac{ \gamma_h\left(a_i\pm z,u_i\pm y;\omega_1,\omega_2\right)}{\gamma_h\left(\pm2z ,\pm 2y;\omega_1,\omega_2\right)}  =
    \prod_{1\leq i < j \leq 6} \gamma_h\left(a_i+a_j,u_i+u_j; \omega_1,\omega_2\right) \:,
\end{align}
with the balancing conditions $\sum_{i=1}^6a_i=\omega_1+\omega_2$, $\sum_{i=1}^6u_i=0$ and $\epsilon(0)=1$ and $\epsilon(y)=2$ for $y>0$.
This integral is an analog of the Nassrallah-Rahman integral evaluation formula. This integral identity was studied in \cite{Gahramanov:2016ilb, Eren:2019ibl, Bozkurt:2020gyy, Mullahasanoglu:2021xyf, Gahramanov:2022jxz} in the context of the Gauge/YBE correspondence\footnote{For details on this correspondence, please refer to the review papers \cite{Gahramanov:2017ysd,Yamazaki:2018xbx,Gahramanov:2022qge}.}.

The limit of the integral when ${r\rightarrow \infty}$ is given by the following identity of the hypergeometric integral \cite{Eren:2019ibl}
\begin{align}
    \sum_{-\infty}^{\infty}\int_{-\infty}^{\infty}\frac{1}{8\pi}\left(z^2+y^2\right)\prod_{i=1}^{6}\boldsymbol{\Gamma}\left(a_i\pm z,u_i \pm y\right) \: dz = \prod_{1\leq i < j \leq 6 } \boldsymbol{\Gamma}\left(a_i+a_j , u_i+ u_j\right)\:.
\end{align}

\subsection{The integral identity II}

The next we consider the following three-dimensional $\mathcal N=2$ supersymmetric duality \cite{Gahramanov:2016wxi}:

\textbf{Theory A.} supersymmetric QCD with $SU(2)$ gauge group and four flavors, chiral multiplets in the fundamental representation of the gauge group and the flavor group

\textbf{Theory B.} no gauge degrees of freedom, with six mesons and a singlet chiral field.

The equality of the lens partition functions for this duality has the form of the following Askey–Wilson type integral identities \cite{Ruijsenaars:cmp}
\begin{align}
     \sum_{y= 0}^{\lfloor \frac{r}{2} \rfloor}\epsilon(y) &\int_{-\infty}^{\infty}\frac{dz}{2r\sqrt{-\omega_1\omega_2}}\prod_{i=1}^{4}\frac{ \gamma_h\left(a_i\pm z,u_i\pm y;\omega_1,\omega_2\right)}{\gamma_h\left(\pm2z ,\pm 2y;\omega_1,\omega_2\right)} \nonumber \\ = &\gamma_h\left(\omega_1+\omega_2-\sum_{i=1}^{4}a_i,-\sum_{i=1}^{4}u_i;\omega_1,\omega_2\right)  
    \prod_{1\leq i < j \leq 4} \gamma_h\left(a_i+a_j,u_i+u_j; \omega_1,\omega_2\right) \label{IntId4}  \:. 
\end{align}
The integral identity (\ref{IntId4}) was studied in \cite{Mullahasanoglu:2023nes, Catak:2024ygo} in the context of the Gauge/YBE correspondence.

The limit of the integral when ${r\rightarrow \infty}$ yields an analogue of Branges-Wilson integral\cite{branges}. 
\begin{align}
    \sum_{-\infty}^{\infty}\int_{-\infty}^{\infty}\frac{1}{8\pi}\left(z^2+y^2\right)\prod_{i=1}^{4}\boldsymbol{\Gamma}\left(a_i\pm z,u_i \pm y\right) \: dz = \frac{\prod_{1\leq i < j \leq 4 } \boldsymbol{\Gamma}\left(a_i+a_j , u_i+ u_j\right)}{\boldsymbol{\Gamma}\left(\sum_{i=1}^{4}a_i,\sum_{i=1}^{4}u_i\right)} \:.
\end{align}
A more general identity, which encompasses our integral identity as a special case was previously presented in \cite{Sarkissian:2020ipg}. Notably, a similar result was also derived in \cite{Neretin_2020}, with a discrepancy in the multiplicative factor.

\subsection{The integral identity III}

Consider the following duality \cite{Gahramanov:2013rda,Gahramanov:2014ona,Bozkurt:2020gyy}:

\textbf{Theory A.} supersymmetric electrodynamics with $U(1)$ gauge symmetry and six chiral multiplets, half of them transforming in the fundamental representation of the gauge group and another half transforming in the anti-fundamental representation

\textbf{Theory B.} no gauge degrees of freedom, nine gauge invariant mesons transforming in the fundamental representation of the flavor group.

The equality of the lens partition functions for this duality has the form of the following hyperbolic hypergeometric integral identity  \cite{Bozkurt:2020gyy}
\begin{align}
\nonumber\sum_{y=0}^{\lfloor\frac{r}{2}\rfloor}\epsilon(y)\ e^{\frac{\pi i}{2}C}  \int_{-\infty}^{\infty} \frac{dz}{r\sqrt{-\omega_1 \omega_2}} \prod_{i=1}^{3} \gamma_h\left( a_i-z,u_i-y;\omega_1,\omega_2\right) \gamma_h\left( b_i+z,v_i+y;\omega_1,\omega_2\right)\\
    = \prod_{i,j=1}^3 \gamma_h\left( a_i+b_j,u_i+v_j;\omega_1,\omega_2\right) \:.
    \label{IntId3x3}
\end{align}
with the balancing conditions $\sum_{i=1}^3 a_i + b_i = \omega_1+\omega_2$ and $\sum_{i=1}^3 u_i + v_i = 0$. The constant in the exponential term is $C = -2y +(u_1+u_2+u_3-v_1-v_2-v_3)$.

The limit of the hyperbolic hypergeometric integral identity ${r\rightarrow \infty}$ is the following
\begin{align}
    \sum_{-\infty}^{\infty}  \int_{-\infty}^{\infty}\frac{dz}{4\pi i}\prod_{i=1}^{3} \boldsymbol{\Gamma}\left( a_i-z,u_1-y\right)\boldsymbol{\Gamma}\left( b_i+z,v_i+y\right)=\prod_{i,j=1}^{3}\boldsymbol{\Gamma}\left( a_i+b_j,u_i+v_j\right)\:.
    \label{IntId3x3E}
\end{align}
The integral identities (\ref{IntId3x3}) and (\ref{IntId3x3E}) were studied in \cite{Bozkurt:2020gyy, Catak:2021coz} in the context of the Gauge/YBE correspondence.

\subsection{The integral identity IV}

Now we consider the following three-dimensional $\mathcal N=2$ supersymmetric duality:

\textbf{Theory A.} supersymmetric QED with two flavors, chiral multiplets in the fundamental representation of the gauge group and the flavor group

\textbf{Theory B.} no gauge degrees of freedom, with four mesons and a singlet chiral field.

The equality of the lens partition functions for this duality has the following form 
\begin{align}
     \sum_{y= 0}^{\lfloor \frac{r}{2} \rfloor}\epsilon(y) &\int_{-\infty}^{\infty}\frac{dz}{r\sqrt{-\omega_1\omega_2}}  e^{\frac{i\pi}{r\omega_1\omega_2}A} \prod_{i=1}^{2} \gamma_h\left(a_i-z,u_i-y;\omega_1,\omega_2\right) \:\: \gamma_h\left(b_i+z,v_i+y;\omega_1,\omega_2\right) \nonumber \\
    = &\gamma_h\left(\omega_1+\omega_2-\sum_{i=1}^2(a_i+b_i),-\sum_{i=1}^2(u_i+v_i);\omega_1,\omega_2\right)\prod_{i,j=1}^{2} \gamma_h\left(a_i+b_j,u_i+v_j;\omega_1,\omega_2\right) \:,
    \label{IntId2x2}
\end{align}
where the exponential factor is
\begin{align}
    \nonumber A=z( a_1+b_1+a_2+b_2 )-\left( a_1a_2-b_1b_2\right)+\omega_1\omega_2(y\left( u_1+v_1+u_2+v_2\right) -\left( u_1u_2-v_1v_2\right))\:.
\end{align}
The integral identity (\ref{IntId2x2}) is studied in \cite{Catak:2024ygo} in the Gauge/YBE correspondence context.

The following hypergeometric integral identity is obtained under the limit ${r\rightarrow \infty}$ of the lens hyperbolic hypergeometric integral identity (\ref{IntId2x2})
\begin{align}
    \sum_{-\infty}^{\infty}\int_{-\infty}^{\infty} \frac{dz}{4\pi} \prod_{i=1}^{2} \boldsymbol{\Gamma}\left( a_i-z,u_i-y\right)\boldsymbol{\Gamma}\left( b_i+z,v_i+y\right)
    = \frac{\prod_{i,j=1}^{2}
    \boldsymbol{\Gamma}\left( a_i+b_j,u_i+v_j\right)}{\boldsymbol{\Gamma}\left(\sum_{i=1}^2a_i+b_i,\sum_{i=1}^2u_i+v_i\right)} \:.
\end{align}

\section{Conclusions}
In this paper, we study the ordinary limits of the lens hyperbolic hypergeometric integral identities, i.e. the limit of integral identities written in terms of Euler's gamma function. From a gauge theory point of view, the lens hyperbolic hypergeometric integral identities express the equality of the partition functions of $\mathcal N = 2$ supersymmetric gauge theories on the three-dimensional lens space $S_b^3/\mathbb Z_r$, and the ordinary limit of the integral identities is a mathematical tool to obtain dual $\mathcal N = (2, 2)$ supersymmetric gauge theories on the two-dimensional sphere $S^2$. There are many hyperbolic hypergeometric integral identities obtained from supersymmetric gauge theories. We plan to use this reduction for other complicated three-dimensional dualities to obtain new ordinary hypergeometric integrals.


\section*{Acknowledgements}
We thank Ilmar Gahramanov for his valuable discussions and further comments on improving the work. We are supported by the Istanbul Integrability and Stringy Topics Initiative (\href{https://istringy.org/}{istringy.org}). Mustafa Mullahasanoglu is supported by the Scientific and Technological Research Council of Turkey (TÜBİTAK) under
the grant numbers 122F451 and 123N952. Ali Mert Yetkin and Reyhan Yumuşak are supported by TÜBİTAK under the grant number 123F436.

	\bibliographystyle{JHEP}
	\bibliography{references}

\providecommand{\href}[2]{#2}\begingroup\raggedright\begin{thebibliography}{10}

\bibitem{Imamura:2012rq}
Y.~Imamura and D.~Yokoyama, {\it {$S^3/Z_n$ partition function and dualities}},
   {\em JHEP} {\bf 11} (2012) 122,
  [\href{http://xxx.lanl.gov/abs/1208.1404}{{\tt arXiv:1208.1404}}].

\bibitem{Imamura:2013qxa}
Y.~Imamura, H.~Matsuno, and D.~Yokoyama, {\it {Factorization of the
  $S^3/\mathbb{Z}_n$ partition function}},  {\em Phys. Rev.} {\bf D89} (2014),
  no.~8 085003, [\href{http://xxx.lanl.gov/abs/1311.2371}{{\tt
  arXiv:1311.2371}}].

\bibitem{Gahramanov:2016ilb}
I.~Gahramanov and A.~P. Kels, {\it {The star-triangle relation, lens partition
  function, and hypergeometric sum/integrals}},
  \href{http://xxx.lanl.gov/abs/1610.0922}{{\tt arXiv:1610.0922}}.

\bibitem{Benini:2011nc}
F.~Benini, T.~Nishioka, and M.~Yamazaki, {\it {4d Index to 3d Index and 2d
  TQFT}},  {\em Phys. Rev.} {\bf D86} (2012) 065015,
  [\href{http://xxx.lanl.gov/abs/1109.0283}{{\tt arXiv:1109.0283}}].

\bibitem{Yamazaki:2013fva}
M.~Yamazaki, {\it {Four-dimensional superconformal index reloaded}},  {\em
  Theor. Math. Phys.} {\bf 174} (2013) 154--166. [Teor. Mat.
  Fiz.174,177(2013)].

\bibitem{Eren:2019ibl}
E.~Eren, I.~Gahramanov, S.~Jafarzade, and G.~Mogol, {\it {Gamma function
  solutions to the star-triangle equation}},  {\em Nucl. Phys. B} {\bf 963}
  (2021) 115283, [\href{http://xxx.lanl.gov/abs/1912.1227}{{\tt
  arXiv:1912.1227}}].

\bibitem{Benini:2012ui}
F.~Benini and S.~Cremonesi, {\it {Partition Functions of ${\mathcal{N}=(2,2)}$
  Gauge Theories on S$^{2}$ and Vortices}},  {\em Commun. Math. Phys.} {\bf
  334} (2015), no.~3 1483--1527, [\href{http://xxx.lanl.gov/abs/1206.2356}{{\tt
  arXiv:1206.2356}}].

\bibitem{Doroud:2012xw}
N.~Doroud, J.~Gomis, B.~Le~Floch, and S.~Lee, {\it {Exact Results in D=2
  Supersymmetric Gauge Theories}},  {\em JHEP} {\bf 05} (2013) 093,
  [\href{http://xxx.lanl.gov/abs/1206.2606}{{\tt arXiv:1206.2606}}].

\bibitem{Gahramanov:2015cva}
I.~Gahramanov and V.~P. Spiridonov, {\it {The star-triangle relation and 3d
  superconformal indices}},  {\em JHEP} {\bf 08} (2015) 040,
  [\href{http://xxx.lanl.gov/abs/1505.0076}{{\tt arXiv:1505.0076}}].

\bibitem{Gahramanov:2016wxi}
I.~Gahramanov and H.~Rosengren, {\it {Basic hypergeometry of supersymmetric
  dualities}},  {\em Nucl. Phys.} {\bf B913} (2016) 747--768,
  [\href{http://xxx.lanl.gov/abs/1606.0818}{{\tt arXiv:1606.0818}}].

\bibitem{Bozkurt:2020gyy}
D.~N. Bozkurt, I.~Gahramanov, and M.~Mullahasanoglu, {\it {Lens partition
  function, pentagon identity, and star-triangle relation}},  {\em Phys. Rev.
  D} {\bf 103} (2021), no.~12 126013,
  [\href{http://xxx.lanl.gov/abs/2009.1419}{{\tt arXiv:2009.1419}}].

\bibitem{Mullahasanoglu:2021xyf}
M.~Mullahasanoglu and N.~Tas, {\it {Lens Partition Functions and Integrability
  Properties}},  \href{http://xxx.lanl.gov/abs/2112.1516}{{\tt
  arXiv:2112.1516}}.

\bibitem{Gahramanov:2022jxz}
I.~Gahramanov, B.~Keskin, D.~Kosva, and M.~Mullahasanoglu, {\it {On Bailey
  pairs for $ \mathcal{N} $ = 2 supersymmetric gauge theories on $
  {S}_b^3/{\mathbb{Z}}_r $}},  {\em JHEP} {\bf 03} (2023) 169,
  [\href{http://xxx.lanl.gov/abs/2210.1145}{{\tt arXiv:2210.1145}}].

\bibitem{Gahramanov:2017ysd}
I.~Gahramanov and S.~Jafarzade, {\it {Integrable lattice spin models from
  supersymmetric dualities}},  {\em Phys. Part. Nucl. Lett.} {\bf 15} (2018),
  no.~6 650--667, [\href{http://xxx.lanl.gov/abs/1712.0965}{{\tt
  arXiv:1712.0965}}].

\bibitem{Yamazaki:2018xbx}
M.~Yamazaki, {\it {Integrability As Duality: The Gauge/YBE Correspondence}},
  {\em Phys. Rept.} {\bf 859} (2020) 1--20,
  [\href{http://xxx.lanl.gov/abs/1808.0437}{{\tt arXiv:1808.0437}}].

\bibitem{Gahramanov:2022qge}
I.~Gahramanov, {\it {Integrability from supersymmetric duality: a short
  review}},  \href{http://xxx.lanl.gov/abs/2201.0035}{{\tt arXiv:2201.0035}}.

\bibitem{Ruijsenaars:cmp}
R.~S.N.M., {\it A generalized hypergeometric function satisfying four analytic
  difference equations of askey–wilson type},  {\em Comm Math Phys} {\bf 206}
  (1999) 639--690.

\bibitem{Mullahasanoglu:2023nes}
M.~Mullahasanoglu, {\it {The star\textendash{}square relation and the
  generalized star\textendash{}triangle relation from 3d supersymmetric
  dualities I}},  {\em Eur. Phys. J. Plus} {\bf 139} (2024), no.~7 643,
  [\href{http://xxx.lanl.gov/abs/2306.1358}{{\tt arXiv:2306.1358}}].

\bibitem{Catak:2024ygo}
E.~Catak and M.~Mullahasanoglu, {\it {Decorating the gauge/YBE
  correspondence}},  {\em Eur. Phys. J. C} {\bf 84} (2024), no.~9 962,
  [\href{http://xxx.lanl.gov/abs/2403.1448}{{\tt arXiv:2403.1448}}].

\bibitem{branges}
D.~Branges, {\it L. tensor product spaces},  {\em Journal Of Mathematical
  Analysis And Applications} {\bf 38} (1972) 109--148.

\bibitem{Sarkissian:2020ipg}
G.~A. Sarkissian and V.~P. Spiridonov, {\it {The Endless Beta Integrals}},
  {\em SIGMA} {\bf 16} (2020) 074,
  [\href{http://xxx.lanl.gov/abs/2005.0105}{{\tt arXiv:2005.0105}}].

\bibitem{Neretin_2020}
Y.~A. Neretin, {\it An analog of the dougall formula and of the de
  branges–wilson integral},  {\em The Ramanujan Journal} {\bf 54} (Mar.,
  2020) 93–106.

\bibitem{Gahramanov:2013rda}
I.~Gahramanov and H.~Rosengren, {\it {A new pentagon identity for the
  tetrahedron index}},  {\em JHEP} {\bf 11} (2013) 128,
  [\href{http://xxx.lanl.gov/abs/1309.2195}{{\tt arXiv:1309.2195}}].

\bibitem{Gahramanov:2014ona}
I.~Gahramanov and H.~Rosengren, {\it {Integral pentagon relations for 3d
  superconformal indices}},  {\em Proc. Symp. Pure Math.93,165} (2016)
  [\href{http://xxx.lanl.gov/abs/1412.2926}{{\tt arXiv:1412.2926}}].

\bibitem{Catak:2021coz}
E.~Catak, I.~Gahramanov, and M.~Mullahasanoglu, {\it {Hyperbolic and
  trigonometric hypergeometric solutions to the star-star equation}},  {\em
  Eur. Phys. J. C} {\bf 82} (2022), no.~9 789,
  [\href{http://xxx.lanl.gov/abs/2107.0688}{{\tt arXiv:2107.0688}}].

\end{thebibliography}\endgroup

\end{document}